\documentstyle [12pt,fleqn]{article}
\begin{document}
\baselineskip=18pt
\begin{titlepage}
\vspace*{-15 pt}
\begin{flushright}
EFI 91-65 \\
November 1991\\
\end{flushright}
\begin{center}
\vspace{.25 in}
{\large \bf The generalized no-ghost theorem for N=2 SUSY critical
strings
\footnote{work supported in part by DOE grant DE-FG02-90ER40560.}}

\vspace{.3 in}
\centerline{Jadwiga Bie\'{n}kowska}
\vspace{.6 cm}
{\em Department of Physics and \\
 Enrico Fermi Institute\\
 the
University of Chicago \\

   Chicago, Il 60637
}
\end{center}
\vspace{.6 cm}
\baselineskip=18pt
\centerline{\bf Abstract}
\vspace{-1 cm}
\begin{quote}
\noindent

We prove the no-ghost theorem for the
N=2 SUSY strings in (2,2) dimensional flat
Minkowski space. We propose a generalization of this theorem for an
arbitrary
geometry of the N=2 SUSY string theory taking advantage of  the  N=4
SCA
generators present in this model. Physical states are found to be the
highest weight states of the N=4 SCA.
\end{quote}
\vspace{1.5 in}
\end{titlepage}
The N=2 SUSY critical  string theory  with (2,2) real signature seems
to be an
interesting model to test duality properties of  string theories
\cite{vafa:1}.
It is particularly simple because the only degree of freedom in the
model is
believed to be a scalar field playing the role of the K\"{a}hler
potential deformations.
Even though there is no reason to suppose that there are other
physical states present
in this model there was  up to date no explicit no-ghost theorem
proven.

In this letter we present such a proof for the flat
Minkowski space with (2,2) real
signature. The proof is a direct generalization of the modern
version of the
no-ghost theorem by Thorn \cite{thorn:1} which uses the BRST quantisation
prescription \cite{batalin}.
We present the line of the proof for the N=2 critical string theory
pointing out the
differences  between the N=0 \cite{thorn:1} and N=2 case.

The N=2 string theory has one advantage over cases with
lower N. It posseses
an additional symmetry generated by the spectral flow
\cite{seiberg:1}. In 4 real
dimensions (critical N=2 strings) the spectral flow generators are
dimension 1 operators and can be
identified  as currents of the local SU(2) (in Euclidean space) or
SU(1,1)
(in Minkowski type space) groups \cite{egu:ta1}. The enlarged set of
generators of the N=2 theory
includes all the N=4 SCA generators \cite{egu:ta2}, with the gauge
symmetry generators being
only the ones belonging to the N=2 SCA subset. Using this larger set
of the operators
we can look at the states in this model as
states from  the Verma module
of the N=4 SCA. We
show that gauging away the N=2 SUSY is enough to prove that the
 only physical states  are  the highest weight states of the N=4
SCA.
In the   Euclidean model the only physical state is the identity, while
in the Minkowski
case there are enough states left in the physical subspace to make
the model
interesting.

The critical N=2 SUSY string lives in   the two
dimensional complex space with
two complex bosons $a^{\mu(*)}_{n}$ and fermions  $b^{\mu(*)}_{n}$
$\mu=1,2$
\cite{ma:mu}. The free field representation of
the N=2 SCA is:
\begin{eqnarray}
L_{n} & = &-\sum_{s} a^{*}_{n-s}\cdot a_{s}
+\sum_{s}(\frac{n}{2}-s)b^{*}_{n-s}\cdot b_{s} \nonumber \\
G_{n}&=&\sqrt{2} \sum_{s} b^{*}_{n-s}\cdot a_{s} \hspace{1 cm}
G^{*}_{n}=\sqrt{2} \sum_{s} b_{n-s}\cdot a^{*}_{s} \nonumber \\
T_{n}&=&-\frac{1}{2} \sum_{s} b^{*}_{n-s}\cdot b_{s}
\label{1}
\end{eqnarray}
where $[a^{\mu}_{r},a^{nu}_{s}]=-r\eta^{\mu \nu}\delta_{r+s}$,
$\{b^{\mu}_{r},b^{nu}_{s}\}=-\eta^{\mu \nu}\delta_{r+s}$. Dots
signify a scalar
product with respect to the $\eta^{\mu \nu}$ Minkowski ( $diag(1,-1)$
) or Euclidean
( $diag(1,1)$)  hermitian metric. The generators obey the N=2 SCA
comutation relations:
\begin{eqnarray}
\left[ L_{n} , L_{m}\right]
&=&(n-m)L_{n+m}+\frac{1}{2}kn(n^{2}-1)\delta_{n+m} \nonumber \\
\{G_{r} , G^{*}_{s}\} &= &
2L_{r+s}+2(r-s)T_{r+s}+\frac{1}{2}k(4r^{2}-1)\delta_{r+s}  \nonumber \\
\left[L_{m} ,G_{r}\right] &= &(\frac{m}{2}-r)G_{m+r} \hspace{1 cm}
\left[L_{m} ,G^{*}_{r}\right]=(\frac{m}{2}-r)G^{*}_{m+r} \nonumber \\
\left[T_{m} ,G_{r}\right]&=&\frac{1}{2}G_{m+r} \hspace{1 cm}
\left[T_{m} ,G^{*}_{r}\right]=-\frac{1}{2}G^{*}_{m+r} \nonumber \\
\left[L_{m} ,T_{n}\right]&=&-nT_{m+r}\hspace{1 cm}
\left[T_{m} ,T_{n}\right]=\frac{1}{2}km\delta_{n+m}
\label{2}
\end{eqnarray}
and all other commutators are zero . The moding of
$b^{\mu(*)}_{r},G^{(*)}_{r}$ operators is governed by the spectral flow
parameter
$\eta$ \cite{seiberg:1},\cite{ma:mu} and we will restrict our attention to  one
sector, let us say the NS with $r\in {\bf Z} + \frac{1}{2}$. The physical
states are
defined by the conditions
\begin{eqnarray}
L_{n}|s\rangle &=&G_{r}|s\rangle=G^{*}_{r}|s\rangle=T_{n}|s\rangle= 0
\,\, for \,\, n,r>0 \,\,\,
L_{0}|s\rangle = T_{0}|s\rangle=0
\label{3}
\end{eqnarray}
In the BRST formulation of the theory for each gauge constraint one
introduces a pair
of ghosts with a statistics  opposite to that of the constraint
\cite{ma:mu},\cite{batalin}. For N=2
there are four pairs of ghosts, associated with
$L_{n},T_{n},G_{n},G^{*}_{n}$
gauge constraints respectively, satisfying the commutation relations
$\{ c_{n}, \bar{c}_{m} \}=\delta_{m+n}$,
$\{ t_{n}, \bar{t}_{m} \}=\delta_{m+n}$, $[ \gamma_{n},
\bar{\gamma}_{m}]=\delta_{m+n}$,
$[ \gamma^{*}_{n}, \bar{\gamma}^{*}_{m}]=\delta_{m+n}$. The BRST
operator $Q$ expressed in terms of these operators is
\begin{eqnarray}
Q &=& L_{n}c_{-n}+G_{p}\gamma_{-p}+G^{*}_{p}\gamma^{*}_{-p}
+T_{n}t_{-n} -\frac{1}{2}(m-n)c_{-m}c_{-n}\bar{c}_{n+m}
-2\gamma_{-p}\gamma^{*}_{-q}\bar{c}_{p+q} \nonumber \\ &-&
2(p-q)\gamma_{-p}\gamma^{*}_{-q}\bar{t}_{p+q}
+  n
c_{-m}t_{-n}\bar{t}_{m+n}+(\frac{m}{2}-p)c_{-m}(\gamma_{-p}\bar{\gamma}_{m+p}+
\gamma^{*}_{-p}\bar{\gamma}^{*}_{m+p})\nonumber \\
&+&\frac{1}{2}t_{-m}(\gamma_{-p}\bar{\gamma}_{m+p}-
\gamma^{*}_{-p}\bar{\gamma}^{*}_{m+p})
\label{4}
\end{eqnarray}
$Q^{2}=0$ for two complex bosons and fermions \cite{ma:mu}.

We are  set to begin  the proof of the no ghost
theorem for the  Minkowski signature  N=2 SUSY critical string theory.
We follow closely the line of the proof proposed by Thorn
\cite{thorn:1}. In  Euclidean space the $L_{0}|state\rangle=0$ condition
trivially forces the only physical state to be the identity.
To fix the gauge conditions (~\ref{3}) we introduce the gauge fixing
operators $A^{+}_{n},B^{+}_{n},A^{*-}_{n},B^{*-}_{n}$ where the notation
$(+,-)$ is
unambiguous in the (1,1) signature  and $A^{\pm (*)}_{n}(B^{\pm (*)}_{r})$
refer to the
free boson  (fermion) operators projected on the lightlike directions.
Following
the line of the proof \cite{thorn:1} we want to construct the Hilbert space of
this
model using  gauge generators, gauge fixing conditions and  states which are
simultanously annihilated by all of them. We consider the states
\begin{eqnarray}
&L^{\lambda_{1}}_{-1}& \ldots L^{\lambda_{l}}_{-l}
G^{g_{1}}_{-1} \ldots G^{g_{r}}_{-r}
G^{*g^{*}_{1}}_{-1} \ldots G^{*g^{*}_{r^{*}}}_{-r^{*}}
T^{\tau_{1}}_{-1} \ldots T^{\tau_{n}}_{-n} \nonumber \\
&A^{+\alpha_{1}}_{-1}& \ldots A^{+\alpha_{a}}_{-a}
B^{+\beta_{1}}_{-1} \ldots B^{+\beta_{b}}_{-b}
A^{*-\alpha^{*}_{1}}_{-1} \ldots A^{*-\alpha^{*}_{a^{*}}}_{-a^{*}}
B^{*-\beta^{*}_{1}}_{-1} \ldots
B^{*-\beta^{*}_{b^{*}}}_{-b^{*}}|t\rangle
\label{5}
\end{eqnarray}
where
$L_{n}|t\rangle=G_{r}|t\rangle=G^{*}_{r}|t\rangle=T_{n}|t\rangle=
A^{*+}_{n}|t\rangle=B^{*+}_{r}|t\rangle=A^{-}_{n}|t\rangle=
B^{-}_{r}|t\rangle=0$ for $n,r>0$ and $L_{0}|t\rangle=h|t\rangle$.
The "mixed" $(+,-)$ gauge fixing conditions are needed to assure the linear
independence of the vectors (~\ref{5}), as we explain below.

The first step   is to prove the linear independence
of the vectors having the form
(~\ref{5}) at each level N. First of all we notice that the number of
generators in the basis (~\ref{5}) is the same as the number of generators in
the
Hilbert space spanned by the free boson and free fermion operators acting on
the
vacuum with a momentum $p$ $|0,p\rangle$, so at each level  the number of
vectors
from the set (~\ref{5}) is the same as the number of  vectors from the free
field
representation of the Hilbert space which clearly forms  a  linearly
independent set
(a basis of the vector space at each mass level).

 What remains to be proven then is that at each
level N we can express any vector from the basis of
$A^{\pm}_{n},A^{*\pm}_{m},B^{\pm}_{r},B^{*\pm}_{s}$ by the vectors from
(~\ref{5}).
This is enough since at each level N we have a finite dimensional vector space.
The
proof follows by induction. For $N=\frac{1}{2}$ the check is trivial since the
$G_{-\frac{1}{2}},G^{*}_{-\frac{1}{2}}$ generators take the role of
$B^{-}_{-\frac{1}{2}},B^{*+}_{-\frac{1}{2}}$ operators. At level $N=1$ we
check by
straightforward algebra  that any vector from the free field basis can be
expressed by
vectors from the set (~\ref{5}). In particular at this level it becames clear
that we
need the mixed gauge fixing conditions as above to ensure the linear
independence of
vectors  (~\ref{5}). We require  that both $A^{\pm}_{0}$ are different from
zero. The
case when either of them is equal zero is easy to prove since the physical
state
condition $L_{0}|s\rangle=0$  automatically requires that there are no  mass
excitations different from $0$. The $p=0$ case has to be addressed separately
and we
discuss it at the end of the main line of the proof.

Having checked the validity of our statement at level
1, we assume that it is true for any vector at level $n<N$, i.e. states from
free field
basis at mass level $n$ can be expressed in terms of vectors (~\ref{5}). Any
state
from the free field basis at level $N$ which is a product of several $A^{\pm
(*)}_{n}(B^{\pm (*)}_{r})$ operators  can be expressed in the basis (~\ref{5})
by the
induction assumption and using the commutation relations with the gauge
generators.
The $B^{*+}_{-N+\frac{1}{2}}|h\rangle,B^{-}_{-N+\frac{1}{2}}|h\rangle $ states
are
expressed easily by
$G_{-N+\frac{1}{2}}|h\rangle=A^{-}_{0}B^{*+}_{-N+\frac{1}{2}}|h\rangle +
other\,\,
states\,\,  from\,\, the\,\, basis$,

$G^{*}_{-N+\frac{1}{2}}|h\rangle=A^{*+}_{0}B^{-}_{-N+\frac{1}{2}}|h\rangle +
other
\,\,states\,\,  from\,\, the\,\, basis$. The states
$A^{*+}_{-N}|h\rangle,A^{-}_{-N}|h\rangle $   can also be expressed in this
basis as
follows. Consider the expansion of the state $G_{-p}G^{*}_{-N+p}|h\rangle$ (
$\frac{1}{2}<p<N-\frac{1}{2}$ and $p\in {\bf Z}+\frac{1}{2}$ ) in the free
fields  basis. We see that it contains the states
$(A^{*-}_{0}A^{+}_{-N}+A^{*+}_{0}A^{-}_{-N})|h\rangle$ plus other states
including
the product of the several operators from the free field  basis. By induction
 we can express these states in the (\ref{5}) basis with one  exception.
There is the possibility that  doing so we will cancel the
$A^{*+}_{0}A^{-}_{-N})|h\rangle$ term. The cancellation occurs from the state
$2A^{-}_{0}A^{*+}_{0}B^{-}_{p-N}B^{*+}_{-p}|h\rangle$ if we  try to
express the $B^{*+}_{-p}$ operator using the $G_{-p}$ gauge generator and then
commute the $B^{-}_{-N+p}$ to the right. To solve this problem we realize that
$T_{-N}=-\frac{1}{2}\sum_{p=\frac{1}{2}}^{N-\frac{1}{2}}(B^{*+}_{-p}B^{-}_{p-N}
+B^{*+}_{-p}B^{-}_{p-N})|h\rangle$ and so the state
$\sum_{p=\frac{1}{2}}^{N-\frac{1}{2}}B^{*+}_{-p}B^{-}_{p-N}$ can be expressed
in the
basis as shown above. Then it follows  from
$(\sum_{p=\frac{1}{2}}^{N-\frac{1}{2}}G_{-p}G^{*}_{-N+p}
+4A^{-}_{0}A^{*+}_{0}T_{-N})|h\rangle= -2NA^{*+}_{0}A^{-}_{-N} +
other\,\,states$
where now all $other\,\, states$ are from the basis (\ref{5}), that we can
express
$A^{-}_{-N}$ in this basis.   The remaining state $A^{*+}_{-N}|h\rangle$ can
 then be expressed by expanding $L_{-N}|h\rangle= A^{-}_{0}A^{*+}_{-N}|h\rangle
+ other \,\,
states$ where all {\em other states} can be expressed in the basis (~\ref{5}).
This
completes the proof of the linear independence of the vectors from the set
(~\ref{5}).

The proof of the no-ghost theorem is now relatively
simple if we follow Thorn's approach \cite{thorn:1}. We consider the  BRST
Hilbert
space built as follows from ghosts and vectors (~\ref{5}). We introduce the N=2
SCA
generators which include the ghosts operators \cite{ma:mu}:
\begin{eqnarray}
\{Q,\bar{c}_{n}\}&=&L_{n}+L_{n}^{gh}={\cal L} _{n} \hspace{1 cm}
\{Q,\bar{t}_{n}\}=T_{n}+T_{n}^{gh}={\cal T} _{n}\nonumber \\
\left[Q,\bar{\gamma}_{p}\right]&=&G_{p}+G_{p}^{gh}={\cal G} _{p}
\hspace{1 cm}
\left[Q,\bar{\gamma}^{*}_{p}\right]=G^{*}_{p}+G^{*gh}_{p}={\cal
G}^{*} _{p}
\label{6}
\end{eqnarray}

We define the ghost vacuum as
$\bar{c}_{0}|0\rangle_{ghost}=\bar{t}_{0}|0\rangle_{ghost}=0$. Then our BRST
basis is
formed by the states
\begin{eqnarray}
&{\cal L}^{\lambda_{1}}_{-1}& \ldots {\cal L}^{\lambda_{l}}_{-l}
\bar{c}^{\bar{\kappa}_{1}}_{-1}\ldots\bar{c}^{\bar{\kappa}_{\bar{k}}}_{-\bar{k}}
{\cal G}^{g_{1}}_{-1} \ldots {\cal G}^{g_{r}}_{-r}
\bar{\gamma}^{\bar{\chi}_{1}}_{-1}\ldots
\bar{\gamma}^{\bar{\chi}_{\bar{s}}}_{-\bar{s}}
\nonumber \\
&{\cal G}^{*g^{*}_{1}}_{-1}& \ldots {\cal
G}^{*g^{*}_{r^{*}}}_{-r^{*}}
\bar{\gamma}^{*\bar{\chi^{*}}_{1}}_{-1}\ldots
\bar{\gamma}^{*\bar{\chi^{*}}_{\bar{s}^{*}}}_{-\bar{s}^{*}}
{\cal T}^{\tau_{1}}_{-1} \ldots {\cal T}^{\tau_{n}}_{-n}
\bar{t}^{\bar{\nu}_{1}}_{-1}\ldots\bar{t}^{\bar{\nu}_{\bar{m}}}_{-\bar{m}}
\nonumber \\
&A^{+\alpha_{1}}_{-1}& \ldots A^{+\alpha_{a}}_{-a}
c^{\kappa_{1}}_{-1}\ldots c^{\kappa_{k}}_{-k}
B^{+\beta_{1}}_{-1} \ldots B^{+\beta_{b}}_{-b}
\gamma^{\chi_{1}}_{-1}\ldots \gamma^{\chi_{s}}_{-s}
\nonumber \\
&A^{*-\alpha^{*}_{1}}_{-1}& \ldots A^{*-\alpha^{*}_{a^{*}}}_{-a^{*}}
t^{\kappa^{*}_{1}}_{-1}\ldots t^{\kappa^{*}_{k^{*}}}_{-k^{*}}
B^{*-\beta^{*}_{1}}_{-1} \ldots B^{*-\beta^{*}_{b^{*}}}_{-b^{*}}
\gamma^{*\chi^{*}_{1}}_{-1}\ldots \gamma^{*\chi^{*}_{s^{*}}}_{-s^{*}}
|t\rangle_{gh}
\label{7}
\end{eqnarray}
where $|t\rangle_{gh}$ means now a tensor product of the ghost vacuum
and previously defined
$|t\rangle$ states from the physical sector.
The above vectors form a linearly independent set  since adding
ghosts does not destroy the linear independence of the basis (~\ref{5}). A more
useful  for the no-ghost theorem is the basis in which we  symmetrise  the
bosonic
operators and antisymmetrise the fermionic ones. Using the N=2 SCA commutation
rules
(~\ref{2}) and the ghosts commutation properties we can do it without spoiling
the
linear independence of the basis (~\ref{7}). We introduce  shorthand notation
to
refer to this symmetrised basis
\begin{eqnarray}
 & {\cal L}^{\lambda}_{\{-l\}}&
\bar{c}^{\bar{\kappa}}_{\left[-\bar{k}\right]}\,\,
{\cal G}^{g}_{\left[-r\right]}\,\,
\bar{\gamma}^{\bar{\chi}}_{\{-\bar{s}\}} \,\,
 {\cal G}^{*g^{*}}_{\left[-r^{*}\right]}\,\,
\bar{\gamma}^{*\bar{\chi^{*}}}_{\{-\bar{s}^{*}\}}\,\,
{\cal T}^{\tau}_{\{-n\}}\,\,
\bar{t}^{\bar{\nu}}_{\left[-\bar{m}\right]}
\nonumber \\
& A^{+\alpha}_{\{-a\}}&
 c^{\kappa}_{\left[-k\right]}\,\,
B^{+\beta}_{\left[-b\right]}\,\,
\gamma^{\chi}_{\{-s\}}\,\,
 A^{*-\alpha^{*}}_{\{-a^{*}\}}\,\,
 t^{\kappa^{*}}_{\left[-k^{*}\right]}\,\,
 B^{*-\beta^{*}}_{\left[-b^{*}\right]}\,\,
\gamma^{*\chi^{*}}_{\{-s^{*}\}}\,\,
|t\rangle_{gh}
\label{8}
\end{eqnarray}
where the $\{\},[\,]$ means symmetrization or antisymmetrization in
the indices enclosed.

Following the proof by Thorn \cite{thorn:1} we can
easily verify that the specific symmetry combination of the first pair ${\cal
L}_{-n}, \bar{c}_{-m}$ can be expressed by other states with a lower number of
$\cal
L$'s and a pure gauge. Working down the number of $\cal L$'s we can express any
state
with a symmetry associated to the Young tableau
\makebox[1in]{
\raisebox{-0.6cm}{$^{1}$} \hspace{-0.5cm}
\raisebox{-0.3cm}{$^{+}$}\hspace{-0.35cm}
$^{\bar{\kappa}}$
 \raisebox{0.22cm}{\fbox{\rule[-0.6cm]{0cm}{0.2cm}\hspace{.2cm}}}
\hspace{-0.7cm}
\raisebox{1.2ex}{\framebox[1.7 cm]{\rule[-0.2cm]{0cm}{0.2cm}\hspace{1
cm}}} \hspace{-1.2 cm} $\mid$\hspace{0.3cm}$\mid$ \hspace{-1cm}
\raisebox{.3cm}{$^{\lambda }$}\hspace{1cm} }
 by other states and a pure gauge . Thus we are left with the states
of the symmetry
\makebox[1 in]{
$^{\bar{\kappa}}$
 \raisebox{0.22cm}{\fbox{\rule[-0.6cm]{0cm}{0.2cm}\hspace{.2cm}}}
\hspace{-0.7cm}
\raisebox{1.2ex}{\framebox[1.7 cm]{\rule[-0.2cm]{0cm}{0.2cm}\hspace{1
cm}}} \hspace{-1.2 cm} $\mid$\hspace{0.3cm}$\mid$ \hspace{-1cm}
\raisebox{.3cm}{$^{\lambda+1}$}\hspace{1cm} }
only. Further requirement that these states be physical ( annihilated
by $Q$) drives us to the conclusion that the symmetry properties exclude any
state
with $(\lambda,\bar{\kappa})\neq(0,0)$  to be physical  (for a more detailed
discussion see \cite{thorn:1}).

This procedure  can be repeated smoothly for all
other pairs of gauge generators and antighosts. We are left with the basis
formed by
the gauge fixing operators and ghosts $ A^{+\alpha}_{\{-a\}}
 c^{\kappa}_{\left[-k\right]}\,\,
B^{+\beta}_{\left[-b\right]}\,\,
\gamma^{\chi}_{\{-s\}}\,\,
 A^{*-\alpha^{*}}_{\{-a^{*}\}}\,\,
 t^{\kappa^{*}}_{\left[-k^{*}\right]}\,\,
 B^{*-\beta^{*}}_{\left[-b^{*}\right]}\,\,
\gamma^{*\chi^{*}}_{\{-s^{*}\}}\,\,
|t\rangle_{gh}$. We can easily see from the commutation relations
\begin{eqnarray}
\left[Q,A^{+}_{-m}\right]&=&m c_{-n}A^{+}_{n-m}
+m\sqrt{2}\gamma^{*}_{-n}B^{+}_{n-m}
 \nonumber \\
\left[Q,A^{*-}_{-m}\right]&=& m c_{-n}A^{*-}_{n-m}
+m\sqrt{2}\gamma_{-n}B^{*-}_{n-m} \nonumber \\
\{Q,B^{+}_{-m}\}&=&( m-\frac{n}{2}) c_{-n}B^{+}_{n-m}
+\frac{1}{2}t_{-n}B^{+}_{n-m} +\sqrt{2}\gamma_{-n}A^{+}_{n-m} \nonumber \\
\{Q,B^{*-}_{-m}\}&=&( m-\frac{n}{2}) c_{-n}B^{*-}_{n-m}
+\frac{1}{2}t_{-n}B^{*-}_{n-m} +\sqrt{2}\gamma_{-n}A^{*-}_{n-m}
\label{9}
\end{eqnarray}
 that the procedure of removing specific symmetry combinations of
gauge fixing conditions and ghosts works when $A^{\pm}_{0}\neq0$. We can
remove,  in
the standard way \cite{thorn:1}, the first pair $A^{+\alpha}_{\{-a\}}
\,\,c^{\kappa}_{\left[-k\right]}$.
Then  by the BRST charge $Q$ acting on the vector  formed from the basis
$B^{+\beta}_{\left[-b\right]}\,\,
\gamma^{\chi}_{\{-s\}}\,\,
B^{*-\beta^{*}}_{\left[-b^{*}\right]}\,\,
\gamma^{*\chi^{*}}_{\{-s^{*}\}}\,\,
 A^{*-\alpha^{*}}_{\{-a^{*}\}}\,\,
 t^{\kappa^{*}}_{\left[-k^{*}\right]}\,\,
 |t\rangle_{gh}$ we will reproduce, due to the commutation relations
(~\ref{9}), the states containing operators $A^{+}_{-n},c_{-m}$\cite{thorn:1}.
By the
previous step, since  we consider the states annihilated by Q,  states
containing any
of these operators  have to be a pure gauge. The procedure of removing the
states
with $(\beta,\chi)\neq(0,0)$ and $(\beta^{*},\chi^{*})\neq(0,0)$  from this
basis
follows.

We are left  with the states formed from the basis
$A^{*-\alpha^{*}}_{\{-a^{*}\}}\,\,
 t^{\kappa^{*}}_{\left[-k^{*}\right]}\,\,
 |t\rangle_{gh}$. We require that for  a linear combination of the
above states $|s\rangle$ $Q|s\rangle=0$. Looking at  the commutation relations
(~\ref{9}) and $[Q,t_{-n}]=m c_{m-n}t_{-m}-2(m-2n)\gamma_{m-n}\gamma^{*}_{-m}$
we see
that it is enough to consider only  the $\gamma^{*}_{-p}$ ghost conservation
number.
The states with the non zero $\gamma^{*}_{-p}$ ghost have to cancel among
themselves
and  starting from the state with the highest $t_{-n}$ excitation we see that
the
only physical state $Q|phys\rangle=0$ is the one with no $t$ ghost.  Then it is
easy
to see that the states built only from $A^{*-}_{-n}$ excitations are not
physical.
This completes the proof for $A^{\pm}\neq0$.

For $A^{+}=0$ or $A^{-}=0$ the physical state
condition $L_{0}|phys\rangle=0$ automatically excludes any excitation so we are
left
with the physical states subspace $|0,p\rangle$ where $p^{\mu}p^{*}_{\mu}=0$.

The $p^{\mu}=0$ case is special, as we know from the
N=0,1 SUSY case \cite{frenkel}. We have to check directly which states with
zero
$L_{0},T_{0}$ eigenvalue are physical. In this case $c_{1}|0\rangle_{gh}\neq0$
$(L_{-1}|0\rangle=0)$,
$\gamma_{\frac{1}{2}}|0\rangle_{gh}\neq0$
$(G_{-\frac{1}{2}}|0\rangle=0)$,
$\gamma^{*}_{\frac{1}{2}}|0\rangle_{gh}\neq0$
$(G^{*}_{-\frac{1}{2}}|0\rangle=0)$ and
there are several nontrivial possibilities  for the states with non
zero ghost numbers to be physical $L_{0}|s\rangle=0$. We investigated all
possible
states built up from the above ghosts and we found that there are in this case
additional
non zero ghost number physical states
\begin{eqnarray}
(a^{\mu}_{-1}c_{1}+\sqrt{2}b^{\mu}_{-\frac{1}{2}}\gamma^{*}_{\frac{1}{2}})
|0,p^{\mu}=0\rangle \hspace{1 cm}
(a^{*\mu}_{-1}c_{1}+\sqrt{2}b^{*\mu}_{-\frac{1}{2}}\gamma_{\frac{1}{2}})
|0,p^{\mu}=0\rangle
\label{10}
\end{eqnarray}

This completes the proof for the no-ghost theorem
for the N=2 SUSY critical strings  in flat space.

We  generalize the no-ghost theorem for the case of
an arbitrary geometry of the N=2 SUSY string theory in (2,2) signature real
dimension
space by taking advantage of the larger set of generators existing in these
theories due
to the spectral flow \cite{seiberg:1}, \cite{egu:ta1}. In the case of the (4,0)
real
signature space the group is enlarged by the spectral flow generators
$T^{+},T^{-}$
which can be identified with the generators of the SU(2) group. The full set of
generators is described by the N=4 SCA operators \cite{egu:ta1},
\cite{egu:ta2}. One
can check that in the free field representation (~\ref{1}) the N=2 SCA algebra
is
completed to the N=4 by adding the generators.
\begin{equation}
T^{+}_{n}=\sum_{s}b^{*1}_{n-s}b^{*2}_{s}\hspace{1 cm}
T^{-}_{n}=-\sum_{s}b^{1}_{n-s}b^{2}_{s}
\label{11}
\end{equation}

It is important to have a hermitian conjugate basis
for the operators $(T^{+}_{n})^{\dagger}=T^{-}_{-n}$ since this unambiguously
defines
the commutation relations in the algebra.
The additional generators of the N=4 SCA and their commutation rules
(non-zero ones) are
\begin{eqnarray}
\left[T^{+}_{n},G^{*}_{r}\right]&=&
\sqrt{2}\sum_{t}(b^{*2}_{n+r-t}a^{*1}_{t}-b^{*1}_{n+r-t}a^{*1}_{t})=B_{n+r}
\nonumber \\
\left[T^{-}_{n},G_{r}\right]&=&
-\sqrt{2}\sum_{t}(b^{2}_{n+r-t}a^{1}_{t}-b^{1}_{n+r-t}a^{1}_{t})=-B^{*}_{n+r}
\nonumber \\
\left[T^{+}_{n},T^{-}_{m}\right]&=&2T_{n+m}
+\frac{1}{2}kn\delta_{n+m}\hspace{1 cm}
\left[T_{n},T^{\pm}_{m}\right]=\pm T^{\pm}_{m+n} \nonumber \\
\left[T^{+}_{m},B^{*}_{r}\right]&=&-G_{r+m}\hspace{1 cm}
\left[T^{-}_{m},B_{r}\right]=G^{*}_{r+m} \nonumber \\
\{B_{r},B^{*}_{s}\}&=&2L_{r+s}+2(r-s)T_{r+s}+\frac{1}{2}k(4r^{2}-1)\delta_{r+s}
\nonumber \\
\left[T_{m},B_{r}\right]&=&\frac{1}{2}B_{r+m} \hspace{1 cm}
\left[T_{m},B^{*}_{r}\right]=-\frac{1}{2}B^{*}_{r+m} \nonumber \\
\{B_{r},G_{s}\}&=&2(r-s)T^{+}_{s+r} \hspace{1 cm}
\{B^{*}_{r},G^{*}_{s}\}=-2(r-s)T^{-}_{r+s}
\label{12}
\end{eqnarray}
which is equivalent to the N=4 SCA with the SU(2) subgroup  discussed
in \cite{egu:ta2} with $G_{r}=\frac{1}{\sqrt{2}}(G^{1}_{r}+
\bar{G}^{2}_{r}),\,\,
G^{*}_{r}=\frac{1}{\sqrt{2}}(G^{2}_{r}+ \bar{G}^{1}_{r}),\,\,
B_{r}=\frac{1}{\sqrt{2}}(-G^{1}_{r}+ \bar{G}^{2}_{r}),\,\,
B^{*}_{r}=\frac{1}{\sqrt{2}}(G^{2}_{r}-\bar{G}^{1}_{r})$. The above  defined
 operators are hermitian in the sense $(B_{r})^{\dagger}=B^{*}_{-r},\,\,
(G_{r})^{\dagger}=G^{*}_{-r}$. It is then straightforward to find the
N=4 SCA in the case
 of Minkowski signature space. The algebra has, as  could be
expected, the SU(1,1) subgroup    and the commutations relations for the
generators
$T^{\pm}_{n}$ and $B^{(*)}_{r}$  defined as in  (~\ref{11},~\ref{12}):
\begin{eqnarray}
\left[T^{+}_{n},T^{-}_{m}\right]&=&-2T_{n+m}+\frac{1}{2}kn\delta_{n+m}
\nonumber \\
\left[T^{+}_{m},B^{*}_{r}\right]&=&G_{r+m}\hspace{1 cm}
\left[T^{-}_{m},B_{r}\right]=-G^{*}_{r+m} \nonumber \\
\{B_{r},B^{*}_{s}\}&=&-(2L_{r+s}+2(r-s)T_{r+s})-\frac{1}{2}k(4r^{2}-1)\delta_{r+s}
\label{13}
\end{eqnarray}
where we have  listed only the commutators which are different from
case in (\ref{12}) .

The algebra we obtain in this case is considerably
different from the Euclidean
space algebra, since it contains as a subgroup the noncompact SU(1,1).
 The SU(1,1) group is the part of the rotation group unbroken by the
 background. Typical physical states of the theory belong to finite
 dimensional representations of SU(1,1).
 Fortunately the
representations of this group were classified some time ago  \cite{bargman}.
There
are three basic types of the SU(1,1) representations classified by the
eigenvalues of
the quadratic Casimir operator $C=-g_{ab}T^{a}_{0}T^{b}_{0}$, where
$g_{ab}=diag(-1,-1,1)$  $C|0,j,m\rangle=-j(j+1)|0,j,m\rangle$, and the discrete
values of the compact direction operator
$T^{3}_{0}|0,j,m\rangle=T_{0}|0,j,m\rangle=m|0,j,m\rangle$. The ground states
of the
N=4 SCA theory (~\ref{13})  belong  to the irreducible representations of the
global
SU(1,1) algebra and we introduced notation $|0,j,m\rangle$  where 0 refers to
the
state zero excitation of the N=4 SCA. There are  following irreducible
(non necessarily unitary) SU(1,1)
representations: continuous ones

1)$C^{0}_{j}$, $j=-\frac{1}{2}+i\kappa$ and $m=0,\pm
1,\pm 2\ldots$

2)$C^{\frac{1}{2}}_{j}$, $j=-\frac{1}{2}+i\kappa$ and
$m=\pm \frac{1}{2},\pm \frac{3}{2}\ldots$

3)$E_{j}$, $j\in {\bf R},j\not \in {\bf Z}, 2j\not \in {\bf Z}$
  $m=0,\pm 1,\pm2,\ldots$ or $m=\pm \frac{1}{2}, \pm \frac{3}{2},\ldots$

 and there is no highest state
anihilated either by $T^{+}_{0}$ or $T^{-}_{0}$;
 infinite discrete representations:

4)$D^{+}_{j}$, $j=-\frac{1}{2}, -1, -\frac{3}{2}
\ldots$ and $m=-j,-j+1,\ldots$

5)$D^{-}_{j}$, $j=-\frac{1}{2}, -1, -\frac{3}{2}
\ldots$ and $m=j,j-1,\ldots$

and  finite discrete representations;

6)$F_{j}$, $j=\frac{1}{2},1,\ldots$ and $m=-j,-j+1,\ldots,j-1,j$

 For the $D^{+}_{j}$ representation there exists a
highest state $|s\rangle^{+}=|0,j-j\rangle$ such that
$T^{-}_{0}|s\rangle^{+}=0$ and
for the $D^{-}_{j}$ representation there is a state
$|s\rangle^{-}=|0,j,j\rangle$
such that $T^{+}_{0}|s\rangle^{-}=0$. These remarks are useful in the
generalization of the no-ghost theorem we present below.

 The space of all states in the theory we consider is
a tensor product of the states of the Verma module of the N=4 SCA,  obeying the
commutation relations (~\ref{2},~\ref{12},~\ref{13}) (SU(1,1) case), and some
other
states  which are  the trivial representation of the N=4 SCA. The goal of our
proof
is to show that requirement that the states obey the N=2 SCA gauge conditions
(~\ref{3}) is enough to gauge away all excitations from the N=4 SCA Verma
module.
This leaves the physical states space of the theory containing only the highest
weight
states of the N=4 SCA.

We consider a state from a Verma module built on a
state belonging to one of the global representations of the SU(1,1) algebra.
The
states built from the ordered product of the N=4 SCA operators acting on the
highest
weight state:
\begin{eqnarray}
&L^{\lambda_{1}}_{-1}& \ldots L^{\lambda_{l}}_{-l}
G^{g_{1}}_{-1} \ldots G^{g_{r}}_{-r}
G^{*g^{*}_{1}}_{-1} \ldots G^{*g^{*}_{r^{*}}}_{-r^{*}}
T^{\tau_{1}}_{-1} \ldots T^{\tau_{n}}_{-n} \nonumber \\
&T^{+\alpha_{1}}_{-1}& \ldots T^{+\alpha_{a}}_{-a}
B^{\beta_{1}}_{-1} \ldots B^{\beta_{b}}_{-b}
T^{*-\alpha^{*}_{1}}_{-1} \ldots T^{-\alpha^{*}_{a^{*}}}_{-a^{*}}
B^{*\beta^{*}_{1}}_{-1} \ldots
B^{*\beta^{*}_{b^{*}}}_{-b^{*}}|0,j,m\rangle
\label{14}
\end{eqnarray}
form a good basis for the Verma module, in the sense that there are no
linear combinations of them equaling zero (of course there are linear
combinations of
states with norm  equal to zero). The ordering of the operators ( last four
$T^{+}_{-n},B_{-p},T^{-}_{-n},B^{*}_{-p}$ ) is important for the no-ghost proof
and
should be set respectively to the representation of the $|0,j,m\rangle$ state.
For
the continuous representations $C^{\frac{1}{2}}_{j},C^{0}_{j},E_{j}$ it does
not
matter, for the $D^{-}_{j}$ representations the positive charge  generators
$T^{+}_{m},B_{n}$ should come to the left as in (~\ref{14}) and for the
$D^{+}_{j}$
representations the negative charge operators $T^{-}_{m},B^{*}_{n}$ come first
from
the left.

The no ghost theorem follows in much
the same way as for  flat space. However there are some differences which we
point
out below. We also restrict our discussion to the case when the ground state
belongs
to $D^{-}_{j}$ or continuous representation, but the $D^{+}_{j}$ case is proven
analogously.

To gauge away the N=2 SUSY  we  use the BRST
prescription with the  BRST charge defined by (~\ref{4}).
The full BRST Hilbert space takes the familiar form
\begin{eqnarray}
 & {\cal L}^{\lambda}_{\{-l\}}&
\bar{c}^{\bar{\kappa}}_{\left[-\bar{k}\right]}\,\,
{\cal G}^{g}_{\left[-r\right]}\,\,
\bar{\gamma}^{\bar{\chi}}_{\{-\bar{s}\}} \,\,
 {\cal G}^{*g^{*}}_{\left[-r^{*}\right]}\,\,
\bar{\gamma}^{*\bar{\chi^{*}}}_{\{-\bar{s}^{*}\}}\,\,
{\cal T}^{\tau}_{\{-n\}}\,\,
\bar{t}^{\bar{\nu}}_{\left[-\bar{m}\right]}
\nonumber \\
& T^{+\alpha}_{\{-a\}}&
 c^{\kappa}_{\left[-k\right]}\,\,
B^{\beta}_{\left[-b\right]}\,\,
\gamma^{\chi}_{\{-s\}}\,\,
 T^{-\alpha^{*}}_{\{-a^{*}\}}\,\,
 t^{\kappa^{*}}_{\left[-k^{*}\right]}\,\,
 B^{*\beta^{*}}_{\left[-b^{*}\right]}\,\,
\gamma^{*\chi^{*}}_{\{-s^{*}\}}\,\,
|0,j,m\rangle_{gh}
\label{15}
\end{eqnarray}

Adding ghosts and the (anti) symmetrization procedure
does not spoil the linear independence of the basis (~\ref{14}). But unlike in
the
previous case the subspace ${\cal H}_{TB}$ built from the vectors  $
T^{+\alpha}_{\{-a\}}
 c^{\kappa}_{\left[-k\right]}\,\,
B^{\beta}_{\left[-b\right]}\,\,
\gamma^{\chi}_{\{-s\}}\,\,
 T^{-\alpha^{*}}_{\{-a^{*}\}}\,\,
 t^{\kappa^{*}}_{\left[-k^{*}\right]}\,\,
 B^{*\beta^{*}}_{\left[-b^{*}\right]}\,\,
\gamma^{*\chi^{*}}_{\{-s^{*}\}}\,\,
|0,j,m\rangle_{gh}$ is not $Q$ invariant. There exist vectors
$|s\rangle \in {\cal H}_{TB}$ such that $Q|s\rangle \not\in {\cal H}_{TB}$.
This is
the consequence of the commutation relations of these generators with the BRST
charge
\begin{eqnarray}
\left[Q,T^{+}_{-m}\right]&=& m c_{-n}T^{+}_{n-m} +T^{+}_{n-m}t_{-n}
-\gamma^{*}_{-n}B_{n-m} \nonumber \\
\left[Q,T^{-}_{-m}\right]&=& m c_{-n}T^{-}_{n-m}-T^{-}_{n-m}t_{-n}
+\gamma_{-n}B^{*}_{n-m} \nonumber \\
\{Q,B_{-m}\}&=&-( m+\frac{n}{2}) B_{n-m}c_{-n}
-\frac{1}{2}B_{n-m}t_{-n} -2(n+m)\gamma_{-n}T^{+}_{n-m} \nonumber \\
\{Q,B^{*}_{-m}\}&=&-( m+\frac{n}{2}) B^{*}_{n-m}c_{-n}
+\frac{1}{2}B^{*-}_{n-m}t_{-n} +(n+m)\gamma_{-n}T^{-}_{n-m}
\label{16}
\end{eqnarray}
and the N=4 SCA (~\ref{12}), (~\ref{13}).

As in the previous case, we would like  to express the
states with a symmetry
\makebox[1 in]{
\raisebox{-0.6cm}{$^{1}$} \hspace{-0.5cm}
\raisebox{-0.3cm}{$^{+}$}\hspace{-0.35cm}
$^{\bar{\kappa}}$
 \raisebox{0.22cm}{\fbox{\rule[-0.6cm]{0cm}{0.2cm}\hspace{.2cm}}}
\hspace{-0.7cm}
\raisebox{1.2ex}{\framebox[1.7 cm]{\rule[-0.2cm]{0cm}{0.2cm}\hspace{1
cm}}} \hspace{-1.2 cm} $\mid$\hspace{0.3cm}$\mid$ \hspace{-1cm}
\raisebox{.3cm}{$^{\lambda }$}\hspace{1cm}}
 in the ${\cal L}_{\{l\}}^{\lambda}$ and
$\bar{c}^{\bar{\kappa}}_{[\bar{k}]}$ operators  successively by the states with
the lower number of ${\cal L}_{-n}$'s and  pure gauge terms  and eventually
eliminate them from the physical states space.
In the case of the basis (~\ref{15}) there will be some states with the same
number $\lambda$ of ${\cal L}_{-n}$ representing the above symmetry state in
our
induction
procedure \cite{thorn:1}, which did not happen previously. The
unwanted ${\cal L}_{-n}$ generators will come from the commutation relations
(~\ref{16}) and a subsequent commutation of the $B,B^{*}$ generators. We are
getting
one ${\cal L}_{-n}$ generator at the expense of  exchanging two $B,B^{*}$ pair
for     the ghost. For each finite N  level space we can subsequently work down
the  number of ${\cal L}_{-n}$-producing generators eliminating them after a
finite number of steps, and so our procedure of eliminating the symmetry
combination
\makebox[1 in]{
\raisebox{-0.6cm}{$^{1}$} \hspace{-0.5cm}
\raisebox{-0.3cm}{$^{+}$}\hspace{-0.35cm}
$^{\bar{\kappa}}$
 \raisebox{0.22cm}{\fbox{\rule[-0.6cm]{0cm}{0.2cm}\hspace{.2cm}}}
\hspace{-0.7cm}
\raisebox{1.2ex}{\framebox[1.7 cm]{\rule[-0.2cm]{0cm}{0.2cm}\hspace{1
cm}}} \hspace{-1.2 cm} $\mid$\hspace{0.3cm}$\mid$ \hspace{-1cm}
\raisebox{.3cm}{$^{\lambda }$}\hspace{1cm}}
 survives.
Then   we consider how the states with the remaining symmetry
\makebox[1 in]{
$^{\bar{\kappa}}$
 \raisebox{0.22cm}{\fbox{\rule[-0.6cm]{0cm}{0.2cm}\hspace{.2cm}}}
\hspace{-0.7cm}
\raisebox{1.2ex}{\framebox[1.7 cm]{\rule[-0.2cm]{0cm}{0.2cm}\hspace{1
cm}}} \hspace{-1.2 cm} $\mid$\hspace{0.3cm}$\mid$ \hspace{-1cm}
\raisebox{.3cm}{$^{\lambda+1}$}\hspace{1cm}}
 can be made physical and we come to the conclusion that the only
possibility is that $(\lambda,\bar{\kappa})=(0,0)$.
Working further with the basis of the states without the $\cal L$'s
or $\bar{c}$'s we would again encounter the $\cal L$ state  reproduced when we
try to
express the combination
\makebox[1 in]{
$^{g}$
 \raisebox{0.22cm}{\fbox{\rule[-0.6cm]{0cm}{0.2cm}\hspace{.2cm}}}
\hspace{-0.7cm}
\raisebox{1.2ex}{\framebox[1.7 cm]{\rule[-0.2cm]{0cm}{0.2cm}\hspace{1
cm}}} \hspace{-1.2 cm} $\mid$\hspace{0.3cm}$\mid$ \hspace{-1cm}
\raisebox{.3cm}{$^{\bar{\chi} +1}$}\hspace{1cm}}
 by other states and pure gauge terms. But by the first step, and physical
state condition, any such state has to be pure gauge and  the procedure follows
all
the way thus taking care of all the states containing the gauge generators and
antighosts.

The next difference showing up in this case is that
the role of the $A^{+}_{0},A^{*-}_{0}$ operators (which are {\bf C}-numbers in
flat
Minkowski space ) is played by the $T^{\pm}_{0}$ as we see from (~\ref{16}).
Since
the ground states in this theory belong to the SU(1,1) infinite
representations, this
step of the proof also survives. To check it  we need the basis ordered as was
discussed before.

Looking at the commutator
\begin{eqnarray}
\left[Q,\right. T^{+(\alpha^{+}+1)}_{-\{a\}}c^{(\kappa-1)}_{-\left[ k
\right]}\left. \right]&=&
\sum_{i}T ^{+\alpha^{+}_{1}}_{\{-1}\ldots\hat{T}
^{+\alpha^{+}_{i}}_{-i}\ldots
T ^{+\alpha^{+}_{a}}_{-a\}}i T^{+}_{0}c_{-i}c^{\kappa-1}_{-\left[ k
\right]} +\nonumber \\
&+& other \,\,\, states
\label{17}
\end{eqnarray}
we see that the state with the symmetry
\makebox[1 in]{
$^{\kappa}$
 \raisebox{0.22cm}{\fbox{\rule[-0.6cm]{0cm}{0.2cm}\hspace{.2cm}}}
\hspace{-0.7cm}
\raisebox{1.2ex}{\framebox[1.7 cm]{\rule[-0.2cm]{0cm}{0.2cm}\hspace{1
cm}}} \hspace{-1.2 cm} $\mid$\hspace{0.3cm}$\mid$ \hspace{-1cm}
\raisebox{.3cm}{$^{\alpha +1}$}\hspace{1cm}}
 can be expressed by other states and a pure gauge if there always
exists a state at zero level $|0,j,m-1\rangle$ such that
$|0,j,m\rangle=T^{+}_{0}|0,j,m-1\rangle$. This statement is always true in the
continuous and $D^{-}_{j}$ representation since we ordered our basis  to assure
that
the states containing $T^{+}_{n},c_{n},B_{m},\gamma_{m}$ operators can be
always
removed.  We can also remove in this way the states built from the
$T^{-}_{n},t_{n},
B^{*}_{m},\gamma^{*}_{m}$ operators  as long as the state $|0,j,m\rangle$ is
different from the highest weight state of the $D^{-}_{j}$ global
representation
$|s\rangle^{-}$. For this state there does not exists  $|\tilde{s}\rangle$ such
that $T^{-}_{0}|\tilde{s}\rangle=|s\rangle^{-}$. We have to consider  states
built on
the highest state $|s\rangle^{-}$ separately. Any other state from the basis
(~\ref{16}),  which has $|0,j,m\rangle$  some other state from the $D^{-}_{j}$
or a
continuous representation  is either nonphysical or a pure gauge.

We have to check directly which linear combinations
of
$T^{-\alpha^{*}}_{\{-a^{*}\}}\,\,
 t^{\kappa^{*}}_{\left[-k^{*}\right]}\,\,
 B^{*\beta^{*}}_{\left[-b^{*}\right]}\,\,
\gamma^{*\chi^{*}}_{\{-s^{*}\}}\,\,
|s\rangle^{-}$ states can be  physical e.i. annihilated by $Q$. When $Q$ acts
on
 the linear product of the above states we can realise   that the terms
$(\ldots)T^{-}_{0}c_{-n}|s\rangle^{-}=(\ldots)c_{-n}|0,j,j-1\rangle$
 have to cancel by themselves, by the
ghost conservation number and orthogonality of the states in the SU(1,1)
representation $D^{-}_{j}$. This excludes the
possibility of the $T^{-}_{-n}$ excitations in the physical state. Let us look
next at the commutators of $Q$ with the remaining generators.
{}From the rules (~\ref{16}), $[Q,t_{-n}]=m
c_{m-n}t_{-m}-2(m-2n)\gamma_{m-n}\gamma^{*}_{-m}$ and
$[Q,\gamma^{*}_{-n}]=(\frac{3}{2}m
-n)c_{-m}\gamma^{*}_{m-n}-\frac{1}{2}t_{-m}\gamma^{*}_{m-n}$ we  see that the
only
terms containing the $\gamma_{-p}$ ghost come from the  commutators with the
$t_{-n}$
ghost and therefore have to cancel by themselves. Starting from the state
containing
$t_{-n}$ with  the highest $n$ we can easily see that the only possibility is
that
there are no $t_{-n}$ ghost excitations at all.
We are finally left with the linear combination of states
$B^{*\beta^{*}}_{\left[-b^{*}\right]}\,\,
\gamma^{*\chi^{*}}_{\{-s^{*}\}}\,\,
|s\rangle^{-}$, and by looking at the $[Q,B^{*}_{-p}]$ commutators, we see the
states of the form
$T^{-}_{0}\gamma^{*}_{-p}|s\rangle^{-}$ have to cancel each other. Starting
from the state with highest excitation number of $B^{*}_{-n}$
and working down the $B^{*}_{-p}$'s number we see that there is no possibility
to
cancel the last one ($[Q,\gamma^{*}_{p}]$ does not contain the $T^{-}_{0}$) and
so
the only solution to the physical state condition is the one where
$\beta^{*}=0$.  It
is straightforward to see that the remaining states built out of the
$\gamma^{*}_{-p}$
ghosts cannot be physical for nonzero  ghost excitation.
For the $|0,j,m\rangle$ belonging to the finite dimension representations
$F_{j}$ we can gauge away any state  from the basis (~\ref{15})
 with $m>-j$ as above.
We have to consider only the linear combination of states
$T^{-\alpha}_{\{-a\}}\,\,
 c^{\kappa}_{\left[-k\right]}\,\,
 B^{\beta}_{\left[-b\right]}\,\,
\gamma^{\chi}_{\{-s\}}\,\,
 T^{-\alpha^{*}}_{\{-a^{*}\}}\,\,
 t^{\kappa^{*}}_{\left[-k^{*}\right]}\,\,
 B^{*\beta^{*}}_{\left[-b^{*}\right]}\,\,
\gamma^{*\chi^{*}}_{\{-s^{*}\}}\,\,
|0,j,-j\rangle$. We require that $Q$ acting on such states is zero.
 From the commutators
(~\ref{16}) we see that states $[\ldots]T^{+}_{0}c_{-n}|0,j,-j\rangle$
have to cancel among themselves and so $\alpha=0$. We can further eliminate
 any
$B_{-p}$ excitations realising that  $[\ldots]T^{+}_{0}c_{-n}|0,j,-j\rangle$
states have to cancel each other. The remaining states
 $T^{-\alpha^{*}}_{\{-a^{*}\}}\,\,
 t^{\kappa^{*}}_{\left[-k^{*}\right]}\,\,
 B^{*\beta^{*}}_{\left[-b^{*}\right]}\,\,
\gamma^{*\chi^{*}}_{\{-s^{*}\}}\,\,
|0,j,-j\rangle$ are gauged away as in equation (~\ref{17}) since we can always
express $|0,j,-j\rangle=T^{-}_{0}|0,j,-j+1\rangle$.
 This completes the proof
that the states from the Verma module of the
N=4 SCA are either nonphysical or pure gauge in the N=2 SUSY theory.
The only possible physical states are the highest weight states of the N=4 SCA
which
fall into the representations of the global SU(1,1) group.

The physical state
constraints (~\ref{3}) could tell us more  about which of these states are
 physical.  We can think of the Hilbert space of this theory as a
 product of the SU(1,1) Kac-Moody algebra
representation and the space which is trivial with respect to it.
 The $L_{0}|state\rangle=0$ condition does not provide
additional information. From the Sugawara construction for the SU(1,1) group
\cite{dixon} and the value of the central charge  c=6 \cite{egu:ta1} we see
that
$L^{SU(1,1)}_{0}|0,j,m\rangle= -j(j+1)|0,j,m\rangle$ and easilly we get
 $-j(j+1)<0$.
 Since
  the SU(1,1) trivial
part of the  space remains undetermined we can not say which
combination of states from the global SU(1,1) representations  are allowed.
 The charge of the state is that of the SU(1,1)
representation and the requirement $T_{0}|state\rangle=0$  means that we should
consider only the SU(1,1) states with zero charge. This excludes  the discrete
 unitary representations $D^{\pm}_{j}$, $C^{\frac{1}{2}}_{j}$, and noninteger
$m$ $E_{j}$ representations since we have  $m\neq0$ in these cases.

In the  proof presented the Verma modules built
from the trivial $(j=0)$ representation of the global SU(1,1) were not
 considered. In
such a case the presented above line of the proof breaks down since both
operators $T^{\pm}_{0}$ annihilate the ground state of the theory.
 This situation is similar to the zero momentum case in Minkowski space but
 here we are unable to determine, from general arguments,
 what is the $L_{0}$ eigenvalue of the
   $|0,j=0,0\rangle$
 state since we can not say much about its SU(1,1) trivial part.

I would like to thank Emil Martinec for bringing this problem to my
 attention and for many enlightening discussions. I thank also Peter Freund
 and Tohru Eguchi for helpful remarks. This work is submitted in partial
 fulfillment of the requirements for a Ph.D. degree in physics at the
 University of Chicago.

\end{document}